\documentstyle[12pt]{article}

\begin{document}
\begin{center}
{\bf\large SPECTRAL LINE SHAPE OF HIGH--FREQUENCY LOCAL 
VIBRATIONS IN ADSORBED MOLECULAR LATTICES}
\end{center}
\begin{center}
{\bf I. V. Kuzmenko and V. M. Rozenbaum}
\end{center}
\begin{center}
{\sl Institute of Surface Chemistry\\ 
National Academy of Sciences of Ukraine\\ 
Prosp. Nauki 31, Kyiv-22, 252022 Ukraine\\
E-mail: tanya@ap3.bitp.kiev.ua}
\end{center}

\begin{abstract}

We consider high--frequency local vibrations anharmonically 
coupled with low--frequency modes in a planar lattice of 
adsorbed molecules. The effect of lateral intermolecular 
interactions on the spectral line shape for local vibrations 
is analyzed in the limit of the high density of adsorbed 
molecules. It is shown that the spectral line positions and 
widths depend on behaviour of low--frequency distribution 
function for a system of adsorbed molecules. The results 
obtained allows the spectral line characteristics of 
the local vibrations for isotopically diluted 
{$^{13}$}C{$^{16}$}O{$_{2}$} molecules in 
the {$^{12}$}C{$^{16}$}O{$_{2}$} monolayer on 
the NaCl(100) surface to be described in agreement with 
the experimentally measured values. 
\end{abstract}

\vspace{5mm}

It is well known that the spectral line broadening 
for high--frequency local vibrations is caused by
anharmonic coupling between high-- and low--frequency 
modes [1--3]. At a sufficiently high frequency of the 
local vibrations, the dephasing model proved fruitful 
for description of vibrational spectra of adsorbed 
molecules [4--6], as it has afforded a number of exact 
solutions [7--9] and it has also allowed consideration of 
degenerate low--frequency deformation vibrations and their 
intrinsic anharmonicity [10]. If the amount of adsorbed 
molecules suffices to form a monolayer, the dephasing model 
should take into account collectivization of vibrational 
molecular modes [11--12]. The dephasing of collectivized 
high--frequency vibrations of adsorbates by low--frequency 
resonance molecular modes not interacting between themselves 
was studied in Ref.~[11]. On the other hand, it is well known 
that forming of the lattices of inclined-oriented adsorbed 
molecules, e.g., CO and CO{$_{2}$} systems on the NaCl(100) 
surface [15,16] causes the essential coupling for 
the low--frequency deformation modes of 
the adsorbates~[12--14,17,18]. 

In the framework of the conventional exchange dephasing 
model, spectral line shift and width for a local vibration
of a single adsorbed molecule are expressed in terms of
the anharmonic coupling of this vibration with a translational 
or orientational molecular vibration with low frequency 
$\omega_\ell$. The last acquires a resonance nature due to 
its interaction with the quasicontinuous spectrum of crystal 
lattice phonons~[6,10]. For a planar molecular lattice only 
the collective low--frequency molecular modes with quasimoment 
$k<\omega_\ell/c_T$ ($c_T$ is the transverse sound velocity) 
are damped via one--phonon emission~[19]. If the density of 
adsorbed molecules is sufficiently high, i.e., provided 
the inequality
$$
\frac{\omega^2_\ell}{4\pi n_a c_T^2}<<1
$$
is valid ($n_a$ is the number of adsorbed molecules per unit 
area), energy of low--frequency vibration of a molecule is 
dissipated through lateral intermolecular interactions and 
interactions of the molecules with substrate are less 
important. To put it differently, the low--frequency 
vibration of a molecule decays by emission of vibrational 
modes of neighbouring molecules.

   In this paper, we investigate dephasing of local 
vibrations for a system of anharmonically coupled 
high-- and low--frequency molecular vibrations. It 
is shown that the experimentally observed spectra of 
the local vibrations in the planar lattices of adsorbed 
molecules can be described when lateral interactions 
between the low--frequency vibrations of the adsorbed 
molecules are considered. If translational symmetry is 
characteristic of the system, the harmonic parts of its 
Hamiltonian can be introduced in a diagonal form with 
respect to the two--dimensional wave vector ${\bf k}$ 
belonging to the first Brillouin zone of the planar 
lattice~[18]. Then we represent total Hamiltonian as 
a sum of harmonic and anharmonic contributions:

\begin{eqnarray}
H     & = & H_{0}+H_{A}, \\
H_{0} & = & \hbar\Omega_{h}\sum_{\bf k}a^{+}_{\bf k}a_{\bf k}+
        \sum_{\bf k}\hbar\omega_{\bf k}b^{+}_{\bf k}b_{\bf k}, 
\\
H_{A} & = & \frac {\hbar\gamma}{N}\sum_{\bf kk'k''}
a^{+}_{\bf k}a_{\bf k'}b^{+}_{\bf k''}b_{\bf k-k'+k''},
\end{eqnarray}
where $a_{\bf k}$ and $a^{+}_{\bf k}$ ($b_{\bf k}$ and 
$b^{+}_{\bf k}$) are annihilation and creation operators 
for the molecular vibrations of the high frequency 
$\Omega_{h}$ (low frequency $\omega_{\bf k}$), $\gamma$
is the biquadratic anharmonicity coefficient, $N$ is the 
number of molecular lattice sites in the main area. Other 
types of anharmonic coupling do not contribute to the spectral 
line broadening [3--5] with the exception of additionally 
renormalizing the biquadratic anharmonicity coefficient~[20,21], 
so that the model in question is widely involved in 
the interpretation of experimental spectra by fitting 
the parameter values.

In the low--temperature limit ($k_BT<<\hbar\omega_{\bf k}$) 
the spectral line for high--frequency molecular vibrations 
is of the Lorentz--like shape with the shift of its maximum 
$\Delta\Omega$ and width $2\Gamma$ are defined by the following 
relationships~[17]:

\begin{equation}
\left(
\begin{array}{c}
\Delta\Omega \\ 2\Gamma
\end{array}
\right) = 
\left(
\begin{array}{c}
\mbox{Re} \\ -2\mbox{Im}
\end{array}
\right)W, 
\end{equation}

\begin{equation}
W=\frac{\gamma}{N}
\sum_{\bf k}{n \left({\omega}_{\bf k} \right)}
\left[
1-\frac{\gamma}{N}\sum_{\bf k'}
\frac{1}{{\omega}_{\bf k}-{\omega}_{\bf k'}+i0}
\right]^{-1},
\end{equation}
where $n(\omega)=\left[e^{\beta \hbar \omega}-1 \right]^{-1}$ 
is the Bose--factor, $\beta=\left(k_BT\right)^{-1}$.

The spectral characteristics described by expressions~(4) 
and (5) are determined by certain function of lateral 
interaction parameters and of the anharmonicity coefficient 
$\gamma$. Assuming that low--frequency molecular band is 
sufficiently narrow, that is, the inequality 
$\beta\hbar\Delta\omega<<1$ is valid, and substituting 
$n(\omega_{\ell})$ for $n(\omega)$ in expression~(5),
the equation considered takes the form: 

\begin{equation}
W=\gamma n\left(\omega_\ell\right)
\int\limits_{-\infty}^{\infty}
\frac{\varrho(\omega)d\omega}
{1-\gamma\int\limits_{-\infty}^{\infty}
\varrho(\omega')d\omega'
\left[ \omega-\omega'+i0 \right]^{-1}},
\end{equation}
where

\begin{equation}
\varrho(\omega)=\frac{1}{N}\sum_{\bf k}
\delta(\omega-\omega_{\bf k})
\end{equation}
is the low--frequency distribution function for a molecular 
lattice. 

To investigate the lateral interaction effect on basis spectral 
characteristics of high--frequency vibrations, assume that 
the anharmonicity coefficient $\gamma$ is sufficiently small. 
Expanding $W$ in powers of $\gamma$ and retaining the terms of 
the orders $\gamma$ and $\gamma^2$, we derive the following 
expressions for spectral line shift and width:

\begin{eqnarray}
\Delta\Omega & = & \gamma n(\omega_{\ell}), \\  
2\Gamma & = & \frac{2\pi\gamma^{2}}{\eta_{\mbox{\small{eff}}}}
n(\omega_{\ell}),
\end{eqnarray}
where

\begin{equation}
\eta_{\mbox{\small{eff}}}=
\left[ 
\int\limits_{-\infty}^{\infty} \varrho^2(\omega) d\omega 
\right]^{-1}.
\end{equation}
Equations~(8) and (9) express the spectral line shift and 
width in terms of the anharmonic coefficient $\gamma$ and 
the parameter $\eta_{\mbox{\small{eff}}}$. The last 
characterizes lateral interactions of low--frequency 
molecular vibrations. Lateral interactions cause the molecular 
vibrations to collectivize, that is, a molecular vibration with 
the frequency ${\omega}_{\ell}$ transforms into a band of 
collective vibrations in an adsorbed molecular lattice 
with nonzero band width $\Delta\omega$. Then assuming that 
the frequency $\omega_\ell$ is located at the centre of 
the vibrational band, the distribution function~(7) can be 
represented as:

\begin{equation}
\varrho(\omega)=
\Biggl\{
\begin{array}{l}
\Delta\omega^{-1} f(z), \ \ \ 
|\omega-\omega_{\ell}|<{\Delta\omega/2} \\
0, \ \ \ |\omega-\omega_{\ell}|>{\Delta\omega/2} 
\end{array}
\end{equation}
$\left( z=(\omega-\omega_\ell)/\Delta\omega \right)$  
with the dimensionless function $f(z)$ determined by 
the dispersion law for the low--frequency molecular 
vibrations. By virtue of the fact that the distribution 
function~(7) is normalized to unity, the following equation 
is valid:

\begin{equation}
\int\limits_{-1/2}^{1/2} f(z) dz = 1. 
\end{equation}

For $\eta_{\mbox{\small{eff}}}$ (10) to be calculated, 
consideration must be given to specific dispersion laws 
for vibrations induced by anisotropic intermolecular 
interactions. However the contribution of lateral 
interactions to spectral characteristics is frequently 
described by a single parameter, the band width $\Delta\omega$ 
for collectivized vibrations of adsorbates, with function 
$f(z)$ assumed to be equal to the step function 
$\theta\left(1-4z^2\right)$. Then Eq.~(10) takes 
the form:

\begin{equation}
\eta^{(0)}_{\mbox{\small{eff}}}={\Delta\omega}.
\end{equation}

Taking account of the distribution function peculiarities 
leads to the change in the parameter $\eta_{\mbox{\small{eff}}}$. 
Since the function $f(z)$ is normalized by Eq.~(12), 
the expression (10) can be rewritten as:

\begin{equation}
\eta_{\mbox{\small{eff}}}=\eta^{(0)}_{\mbox{\small{eff}}}
\left[ 
1+\int\limits_{-1/2}^{1/2} \left(f(z)-1\right)^2 dz 
\right]^{-1}.
\end{equation}
Eq.~(14) evidently demonstrate the decrease in the parameter 
$\eta_{\mbox{\small{eff}}}$ if the distribution function differs 
from the step function.

To estimate the parameter $\eta_{\mbox{\small{eff}}}$ let us 
assume the circular first Brillouin zone, 
$0<k<k_{\mbox{\small{max}}}$ 
($k_{\mbox{\small{max}}}=\sqrt{4 \pi n_a}$)
and approximate the dispersion law for low--frequency vibrations 
by the following expression:

\begin{equation}
{\omega}_{\bf k}=\omega_\ell-\frac{\Delta\omega}{2}
\left[
1-2\left(1+a\right)\frac{k}{k_{{\mbox{\small{max}}}}}+
2a\left(\frac{k}{k_{{\mbox{\small{max}}}}}\right)^2
\right],  
\end{equation}
with the arbitrary parameter $a$ ($|a|<1$). The relationship~(15) 
involves the linear in $k$ term, which is in agreement with 
the dispersion laws for two--dimensional lattice systems 
constituted by dipole moments~[14]. Substituting 
the expression~(15) into Eq.~(7) and integrating derived 
relation with respect to wave vector ${\bf k}$ we obtain 
the dimensionless distribution function:

\begin{equation}
f(z)=\frac{1}{a}
\left[
\frac{1+a}{\left[\left(1+a\right)^2-2a(2z+1)\right]^{1/2}}-1
\right], \ \ \ |z|<\frac{1}{2}.
\end{equation}
At $a=-1$ the function~(16) reduces to the step function 
$\theta\left(1-4z^2\right)$. Substitution of the expression~(16) 
into Eq.~(10) gives:

\begin{equation}
\eta_{\mbox{\small{eff}}}=\Delta\omega a^2
\left[
\frac{1+a}{2a}\ln\frac{1+a}{1-a}-1-2a
\right]^{-1}.
\end{equation}
Parameter $\eta_{\mbox{\small{eff}}}$ varies from $\Delta\omega$ 
at $a=-1$ to zero at $a=1$. At $a=0.8$ (which is consistent with 
the realistic dispersion laws): 

\begin{equation}
\eta_{\mbox{\small{eff}}}=0.34\Delta\omega. 
\end{equation}

In what follows we consider relationship~(6) (which is true for 
arbitrary values of $\gamma$ and $\eta_{\mbox{\small{eff}}}$), 
with the density of states for the low--frequency band 
approximated by the step function with the step width 
$\eta_{\mbox{\small{eff}}}$. Then Eq.~(6) takes the form:

\begin{equation}
W=\gamma n\left(\omega_\ell\right)
\int\limits_{-1/2}^{1/2}dz
\left[
{1-\frac{\gamma}{\eta_{\mbox{\small{eff}}}}
\ln\frac{1+2z}{1-2z}+
\frac{i\pi\gamma}{\eta_{\mbox{\small{eff}}}}}
\right]^{-1}.
\end{equation}
The expression~(19) allows the lateral interaction parameters 
to be determined by a comparison between observed and calculated 
spectral characteristics.

An example of adsorbed lattices with strong lateral interactions 
is the system of isotopically diluted {$^{13}$}C{$^{16}$}O{$_{2}$} 
molecules in the {$^{12}$}C{$^{16}$}O{$_{2}$} monolayer on 
the NaCl(100) surface. Stretch vibration frequencies for 
{$^{13}$}C{$^{16}$}O{$_{2}$} and {$^{12}$}C{$^{16}$}O{$_{2}$} 
molecules differ by 60~cm{$^{-1}$} and their coupling can thus 
be neglected. In contrast, low--frequency vibrations prove 
to be essentially coupled. As far as translational vibrations 
are concerned, the mass difference between carbon isotopes is 
slight compared to the total molecular mass, whereas for 
orientational vibrations, the mass of the central carbon atom 
has only a slight effect on the molecular moment of inertia. 
Temperature dependent shift and width of spectral line of 
stretch vibrations with the frequency $\Omega_h=2281.6$~cm{$^{-1}$} 
at $T=1$~K [22] can be approximated in the range from $T=1$~K to 
80~K by the following expressions:

\begin{eqnarray}
\Delta\Omega & = & 0.52 n(\omega_{\ell}) \ \mbox{cm}^{-1},
\\
2\Gamma      & = & 0.61 n(\omega_{\ell}) \ \mbox{cm}^{-1},
\end{eqnarray}
and $\omega_\ell=41 \ \mbox{cm}^{-1}$. 
To explain observed spectral characteristics we equate 
expressions (20) and (21) with the real and imaginary parts of 
the $W$ (19) and obtain two equations in two parameters: $\gamma$ 
and $\eta_{\mbox{\small{eff}}}$. Solving this system of equations 
we have the following parameter values: 
$\gamma=0.78\ \mbox{cm}^{-1}$ and 
$\eta_{\mbox{\small{eff}}}=3.8\ \mbox{cm}^{-1}$.
Using Eq.~(18) we obtain the value of effective band width for 
the low--frequency vibrations: $\Delta\omega=11\ \mbox{cm}^{-1}$,
which is in good agreement with the band width for collectivized 
orientational vibrations of CO{$_2$} molecules with 
quadrupole--quadrupole intermolecular interactions. 

We note in conclusion that attempts to describe the 
temperature dependences of the spectral line shape 
for the above-mentioned CO{$_{2}$} molecular ensemble 
in the framework of the conventional exchange dephasing 
model for a single adsorbed molecule lead to overestimated 
spectral line width, whereas calculated line shift agrees 
well with experimental values. Taking into consideration 
the lateral interactions between the low--frequency 
vibrations of the adsorbates, the model advanced provides 
agreement between calculated and observed spectral line 
shift and width.

This research was supported by the State Foundation of 
Fundamental Researches, administered by the State Committee 
on Science and Technology of Ukraine (Project 2.4/308). 
V.M.R. acknowledges also the support of the International 
Science-Education Program (Grant No QSU 082166).

\end{document}